\documentclass[preprint,12pt]{elsarticle}

\usepackage{amssymb}

\newcounter{bla}

\journal{Computer Physics Communications}

\begin{document}

\begin{frontmatter}



\title{PACIAE 2.1: An updated issue of the parton and hadron cascade model       PACIAE 2.0}


\author[a,b]{Ben-Hao Sa\corref{author}}
\author[b]{Dai-Mei Zhou}
\author[a]{Yu-Liang Yan}
\author[a]{Bao-Guo Dong}
\author[b]{Xu Cai}
\cortext[author] {Corresponding author.\\\textit{E-mail address:
sabh@ciae.ac.cn}}
\address[a]{China Institute of Atomic Energy, P. O. Box 275 (10), 102413
Beijing, China.}
\address[b]{Institute of Particle Physics, Central China Normal
University, 430082 Wuhan, China \\ and Key Laboratory of Quark and Lepton
Physics (CCNU), Ministry of Education, China.}

\begin{abstract}
We have updated the parton and hadron cascade model PACIAE 2.0 (cf. Comput.
Phys. Comm. 183 (2012) 333) to the new issue of PACIAE 2.1. The PACIAE model
is based on PYTHIA. In the PYTHIA model, once the generated particle/parton
transverse momentum $p_T$ is randomly sampled, the $p_x$ and $p_y$ components
are originally put on the circle with radius $p_T$ randomly. Now it is
put on the circumference of ellipse with half major and minor axes of
$p_T(1+\delta_p)$ and $p_T(1-\delta_p)$, respectively, in order to better
investigate the final state transverse momentum anisotropy.

\end{abstract}


\end{frontmatter}



{\bf PROGRAM SUMMARY}

\begin{small}
\noindent
{\em Manuscript Title: PACIAE 2.1: An updated issue of the parton and hadron
 cascade model PACIAE 2.0}                                     \\
{\em Authors: Ben-Hao Sa, Dai-Mei Zhou, Yu-Liang Yan, Bao-Guo Dong,
 and Xu Cai} \\
{\em Program Title: PACIAE version 2.1}                       \\
{\em Journal Reference:}      \\
{\em Catalogue identifier:}                                   \\
{\em Licensing provisions: none}                                   \\
{\em Programming language: FORTRAN 77 or GFORTRAN}                        \\
{\em Computer: DELL Studio XPS and others with a FORTRAN 77 or GFORTRAN
 compiler}                                               \\
{\em Operating system: Linux or Windows with FORTRAN 77 or GFORTRAN
 compiler}                                       \\
{\em RAM:} $\approx$ 1G bytes                                 \\
{\em Number of processors used:}                              \\
{\em Supplementary material:}                                 
{\em Classification: 11.1, 17.8}                                         \\
{\em External routines/libraries:}                            \\
{\em Subprograms used:}                                       \\
{\em Catalogue identifier of previous version: aeki\_v1\_0}*              \\
{\em Journal reference of previous version: Comput. Phys. Comm. 183(2012)333.}
\\
{\em Does the new version supersede the previous version?: Yes}*   \\
{\em Nature of problem: PACIAE is based on PYTHIA. In the PYTHIA model, once
 the generated particle/parton transverse momentum ($p_T$) is randomly
 sampled, the $p_x$ and $p_y$ components are randomly placed on the circle
 with radius of $p_T$. This strongly cancels the final state transverse
 momentum asymmetry developed from the initial spatial asymmetry.}\\
{\em Solution method: The $p_x$ and $p_y$ component is now randomly placed on
 the circumference of ellipse with half major axis of $p_T(1+\delta_p)$ and
 the half minor axis of $p_T(1-\delta_p)$ instead of circle.}\\
{\em Reasons for the new version: PACIAE is based on PYTHIA, where once the
 generated particle/parton transverse momentum ($p_T$) is randomly sampled,
 the $p_x$ and $p_y$ components are randomly placed on the circle with radius
 of $p_T$. This is not only strongly canceling the final state transverse
 momentum asymmetry developed from the initial state spatial asymmetry but
 also inconsistent with the ATLAS observation of the final state charged
 particle transverse sphericity is less than unity \cite{atlas}.}\\
{\em Summary of revisions: The main revision is executed by randomly placing
 $p_x$ and $p_y$ components of the generated particle/parton transverse
 momentum $p_T$ on the circumference of ellipse with half major axis of
 $p_T(1+\delta_p)$ and the half minor axis of $p_T(1-\delta_p)$ instead of
 circle.}\\
{\em Restrictions: Depend on the problem studied.}\\
{\em Unusual features:}\\
{\em Additional comments: Email addresses: zhoudm@phy.ccnu.edu.cn (D.-M. Zhou),
 yanyl@ciae.ac.cn (Y.-L. Yan).}\\
{\em Running time:
\begin{itemize}
\item Using the attached input file of usux.dat (where the string
fragmentation is selected and the elastic parton-parton interactions is
considered only, the same later) to run 1000 events for the $\sqrt s$=200 GeV
Non Single Diffractive pp collision by 21a.tar.gz spends 0.5 minute.
\item Using the attached input file of usu.dat to run 10 events for the
10-40\% most central Au+Au collisions at $\sqrt{s_{NN}}$=200 GeV by 21b.tar.gz
spends 5 minutes.
\item Using the attached input file of usu.dat to run 10 events for the
10-40\% most central Au+Au collisions at $\sqrt{s_{NN}}$=200 GeV by 21c.tar.gz
spends 17 minutes.
\end{itemize}
}

\end{small}

\section{}
\label{}



The large azimuthal anisotropy (the large second harmonic coefficient $v_2$)
of the emitted particle is an important feature of the hot and dense medium
created in the ultra-relativistic nuclear collisions. This large $v_2$ has
contributed to the observation of a strongly coupled quark-gluon plasma
(sQGP) in the nucleus-nucleus collisions at the RHIC energies
\cite{brah,phob,star,phen}.

The nuclear overlap zone created in a nucleus-nucleus collisions at a given
impact parameter possesses an almond-like spatial asymmetry. Because of
the strong parton rescattering, the local thermal equilibrium and asymmetric
pressure gradient may build up in this initial fireball. The asymmetric
pressure gradient then drives a collective anisotropic expansion. The
expansion along the almond minor axis (along the large pressure gradient) is
faster than the one along the major axis. This results in a strong asymmetric
transverse momentum azimuthal distribution and hence a large elliptic flow
coefficient $v_2$ of the final hadronic state.

\begin{figure}
\centering
\includegraphics[width=3.0in]{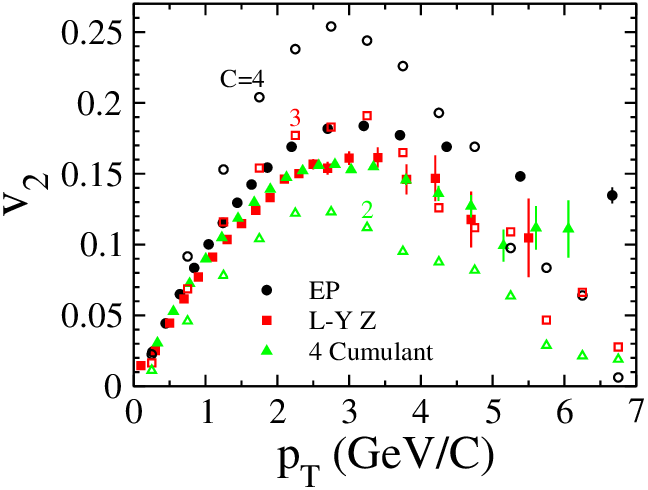}
\caption{(Color on line) The charged particle $v_2(p_T)$ at mid-rapidity
          ($|\eta|<1.$) in 10-40\% most central Au+Au collisions at $\sqrt{s_
          {NN}}$=200 GeV. The solid symbols are the STAR data taken from
          \cite{star1} and open symbols are the PACIAE results calculated
          with $C$=2, 3, and 4.}
\label{v2pt}
\end{figure}

As mentioned in \cite{sa}, PACIAE is a parton and hadron cascade model for the
ultra-relativistic nuclear collisions and is based on PYTHIA \cite{sjo}. In
the PACIAE model, a nucleus-nucleus collision is decomposed into a sequence
of nucleon-nucleon (NN) collisions according to the collision geometry and
the NN total cross section. Each NN collision is performed, in turn, by the
PYTHIA model with the string fragmentation switched-off temporarily and the
diquark (anti-diquark) broken into quark pairs (anti-quark pairs) randomly.
The parton rescattering then proceeds. This parton evolution stage is followed
by the hadronization at the moment of partonic freeze-out (exhausting the
partonic collisions). The Lund string fragmentation regime and/or
phenomenological coalescence model is provided for the hadronization. Then
the rescattering among produced hadrons is dealt with the usual two body
collision model \cite{sa}.

In the PYTHIA model \cite{sjo} once the transverse momentum $p_T$ of a final
state hadron generated from the string fragmentation and/or the unstable
particle decay is randomly sampled, the $p_x$ and $p_y$ components are
randomly placed on the circle with radius of $p_T$. This $p_x$ and $p_y$
determination method may completely cancel the final state transverse momentum
anisotropy developed from the initial spatial asymmetry. The charged particle
transverse sphericity \cite{atlas,cms,alice} may equal to one (isotropic).
This is inconsistent with the experimental observation that the charged
particle transverse sphericity is less than unity \cite{atlas}. Therefore we
are randomly placed the generated final state hadrons on the circumference of
ellipse with half major axis of $p_T(1+\delta_p)$ and the half minor axis of
$p_T(1-\delta_p)$ instead of circle now. This change is also introduced in the
particle/parton production process of hard scattering, multiple interactions,
initial- and final-state parton showers, as well as the adding of remnants
\cite{sjo}. This change is even introduced in the deexcitation of energetic
quark (anti-quark) when the phenomenological coalescence model \cite{sa} is
selected for hadronization. Of course, a new $p_T$ should be recalculated by
$p_x$ and $p_y$ after this change. Then the transverse momentum distribution
of final state hadron may be modified. However, if the deformation parameter
$\delta_p$ is less than unity (a small perturbation) the change in transverse
momentum distribution may be weak.

\begin{figure}
\centering
\includegraphics[width=3.0in]{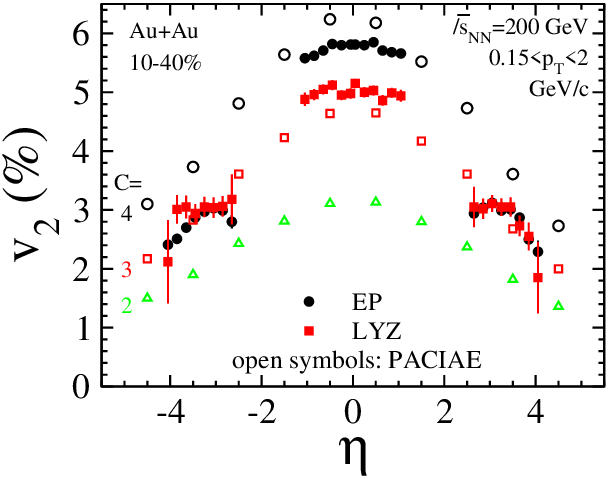}
\caption{(Color on line) The charged particle $v_2(\eta)$ ($0.15<p_T<2.$) in
          10-40\% most central Au+Au collisions at $\sqrt{s_{NN}}$=200 GeV.
          The solid symbols are the STAR data taken from \cite{star1} and
          open symbols are the PACIAE results calculated with $C$=2, 3, and
          4.}
\label{v2eta}
\end{figure}

From ideal hydrodynamic calculations \cite{hein} one knows that the
integrated elliptic flow parameter is directly proportional to the initial
spatial eccentricity of the nuclear overlap zone. Therefore, if the nuclear
overlap zone is assumed to be an ellipse with major axis of
$b=2R_A(1+\delta_r)$ and the minor axis of $a=2R_A(1-\delta_r)$ for a symmetry
nucleus-nucleus collision with the nuclear radius of $R_A$, we may assume
\begin{equation}
\delta_p=C\delta_r
\label{delt}
\end{equation}
where $C$ is an extra model parameter. $C=0$ corresponds to the original case
of $p_x$ and $p_y$ put on the circle randomly.

In order to calculate $\delta_r$ we first calculate the reaction plane
eccentricity \cite{phob1}
\begin{equation}
\epsilon_{rp}=\frac{\sigma_y^2-\sigma_x^2}{\sigma_y^2+\sigma_x^2}
\label{eccc}
\end{equation}
according to the participant nucleons spatial distributions inside the nuclear
overlap zone \cite{sa} in the PACIAE simulation. In the above equation,
$\sigma_x^2=\overline{x^2}-\bar{x}^2$ (the same for $\sigma_y^2$) and
$\bar x^2$ ($\bar x)$ denotes an average of $x^2$ ($x$) over particles in a
single event. The event average reaction plane eccentricity reads
\begin{equation}
\langle\epsilon_{rp}\rangle=\langle\frac{\sigma_y^2-\sigma_x^2}
                                        {\sigma_y^2+\sigma_x^2}\rangle.
\end{equation}
On the other hand, the geometric eccentricity \cite{beye} of the ellipse-like
nuclear overlap zone is
\begin{equation}
\epsilon_g=\sqrt{\frac{b^2-a^2}{b^2}}.
\label{eccg}
\end{equation}
Letting $\epsilon_g=\epsilon_{rp}$, one approximately obtains
\begin{equation}
\delta_r\simeq\frac{\epsilon_{rp}^2}{4}.
\end{equation}

\begin{figure}
\centering
\includegraphics[width=5in]{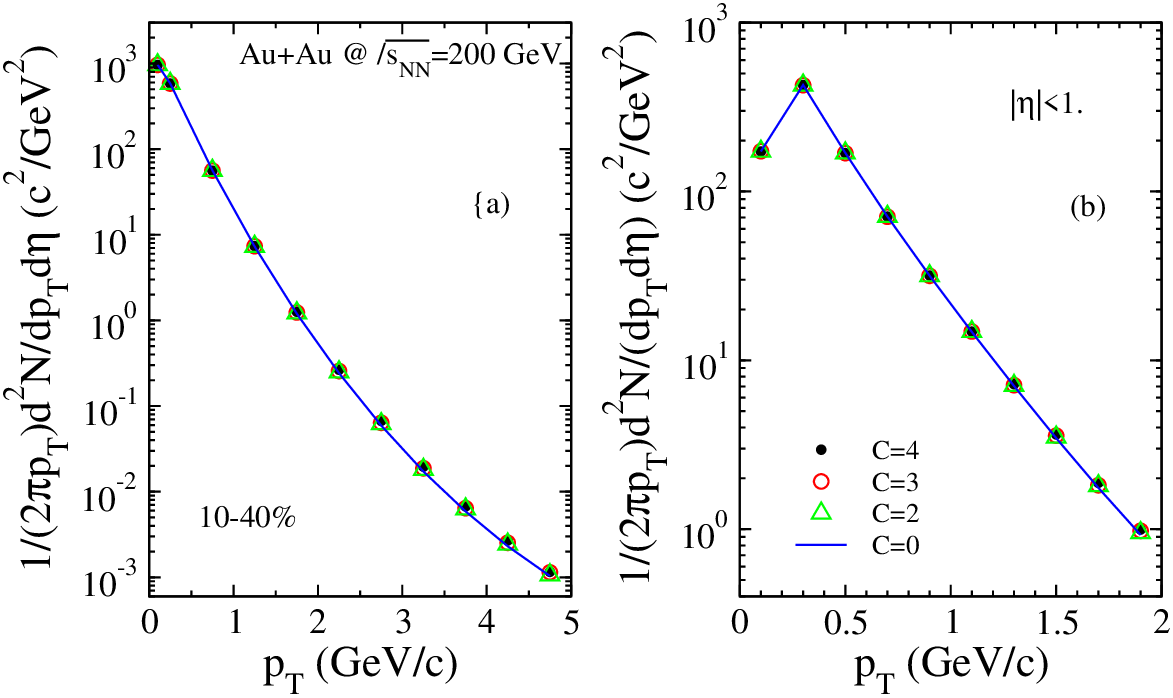}
\caption{(Color on line) The charged particle transverse momentum
          distribution in 10-40\% most central Au+Au collisions at $\sqrt{s_
          {NN}}$=200 GeV calculated by the PACIAE model with $C$=0, 2, 3, and
          4: (a) in the full $\eta$ phase space, (b) in $|\eta|<1$. }
\label{pt}
\end{figure}

The calculated charged particle $v_2(p_T)$ at mid-rapidity ($|\eta|<1.$) in
the 10-40\% most central Au+Au collision at $\sqrt{s_{NN}}$=200 GeV is
compared with the corresponding STAR data \cite{star1} in Fig.~\ref{v2pt}.
In this figure the STAR data are denoted by solid symbols: the black circles
are measured with the event plane method (EP), red squares with Lee-Yang
zero point method (L-YZ), and green triangles with four particle cumulant
method (4 cumulant). The PACIAE results are given by open symbols: the black
circles calculated with $C$=4, red squares with $C$=3, and green triangles
with $C$=2. One sees in this figure that the STAR data \cite{star1} on the
charged particle $v_2(p_T)$ are able to be reproduced by the PACIAE
calculations with $C$=3. The $C$=0 PACIAE results are too small compared to
the STAR data.

A similar comparison for charged particle $v_2(\eta)$ ($0.15<p_T<2.$ GeV/c)
in the 10-40\% most central Au+Au collision at $\sqrt{s_{NN}}$=200 GeV is
given in Fig.~\ref{v2eta}. Here one sees again that the STAR data on the
charged particle $v_2(\eta)$ \cite{star1} are able to be reproduced by the
PACIAE calculations with $C$=3. We have to mention here that in Fig. 3 of
Ref.~\cite{star1} the PHOBOS data \cite{phob2} were introduced to compare
with the STAR data and to complement the lack of the STAR data in $1.5<|\eta|
<2.5$ region. Because the PHOBOS data were measured for the 0-40\% most
central Au+Au collisions at the same energy but in the full $p_T$ phase space,
it is not suitable to compare the PHOBOS data with the STAR data. Therefore
we do not include the PHOBOS data \cite{phob2} in Fig.~\ref{v2eta} here.

We give the calculated charged particle transverse momentum distribution in
10-40\% most central Au+Au collisions at $\sqrt{s_{NN}}$=200 GeV in
Fig.~\ref{pt}. Figure \ref{pt} (a) and (b) are drawn for the full and partial
($|\eta|<1.$) pseudo-rapidity phase space, respectively. In panels (a) and
(b) the solid black circles, open red circles, open green triangles, and the
blue line are calculated with $C$=4, 3, 2, and 0, respectively. We see in this
figure that the charged particle transverse momentum distribution is really
not sensitive to the parameter $C$ both in the full and partial $\eta$ phase
space, provided the deformation parameter $\delta_p$ is less than unity (a
small perturbation).

In addition, an extra switching parameter of $iparres$ is introduced in the
new issue of PACIAE 2.1. We assume that the $iparres$=0 is for the elastic
parton-parton rescattering only and $iparres$=1 for otherwise.


\bibliographystyle{elsarticle-num}
\bibliography{<your-bib-database>}

\begin{thebibliography}{}
\bibitem{atlas}
ATLAS Collaboration, arXiv: 1206.2135v1.
\bibitem{brah}
I. Arsene, et al., BRAHMS Collaboration, Nucl. Phys. A {\bf 757}, 1 (2005).
\bibitem{phob}
B. B. Back, et al., PHOBOS Collaboration, Nucl. Phys. A {\bf 757}, 28 (2005).
\bibitem{star}
J. Admas, et al., STAR Collaboration, Nucl. Phys. A {\bf 757}, 102 (2005).
\bibitem{phen}
K. Adcox, et al., PHENIX Collaboration, Nucl. Phys. A {\bf 757}, 184 (2005).
\bibitem{sa}
Ben-Hao Sa, Dai-Mei Zhou, Yu-Liang Yan, Xiao-Mei Li, Sheng-Qin Feng,
Bao-Guo Dong, and Xu Cai, Comput. Phys. Comm. {\bf 183}, 333 (2012).
\bibitem{sjo}
T. Sj\"{o}strand, S. Mrenna, and P. Skands, JHEP {\bf 05}, 026 (2006).
\bibitem{cms}
CMS Collaboration, Phys. Lett. B {\bf 699}, 48 (2011).
\bibitem{alice}
ALICE Collaboration, arXiv: 1205.3963v1.
\bibitem{hein}
P. F. Kolb, J. Sollfrank, and U. W. Heinz, Phys. Rev. C {\bf 62}, 054909
(2000).
\bibitem{phob1}
Dai-Mei Zhou, Yu-Liang Yan, Bao-Guo Dong, Xiao-Mei Li, Du-Juan Wang, Xu Cai,
and Ben-Hao Sa, Nucl. Phys. A {\bf 860}, 68 (2011); B. Alver, et al., Phys.
Rev. C {\bf 77}, 014906 (2008).
\bibitem{beye}
W. H. Beyer, ``Standard Mathematical Tables and Formulae", p. 177, 29th
Edition, CRC Press, London, 2000.
\bibitem{star1}
B. I. Abelev, et al., STAR Collaboration, Phys. Rev. C {\bf 77}, 054901 (2008).
\bibitem{phob2}
B. B. Back, et al., PHOBOS Collaboration, Phys. Rev. Lett. {\bf 94}, 112303
(2005).

\end{thebibliography}



\end{document}